\pdfoutput=1
\documentclass{article}      

\usepackage{graphicx}
\usepackage{latexsym}
\usepackage{amsmath, amsfonts, amssymb, amsthm}
\usepackage{algorithm}
\usepackage[noend]{algorithmic}
\usepackage{url}
\usepackage{graphicx}
\usepackage[usenames]{color}
\usepackage{subfig}
\usepackage{rotating}
\usepackage{array}
\usepackage{natbib}

\theoremstyle{plain}
\newtheorem{theorem}{Theorem}

\theoremstyle{definition}
\newtheorem{definition}{Definition}

\newcommand{\NNF}{{\tt NNF}}
\newcommand{\dNNF}{{\tt d-NNF}}
\newcommand{\sNNF}{{\tt s-NNF}}
\newcommand{\DNF}{{\tt DNF}}
\newcommand{\dDNF}{{\tt d-DNF}}
\newcommand{\sDNF}{{\tt s-DNF}}
\newcommand{\sdDNF}{{\tt sd-DNF}}

\newcommand{\DNNF}{{\tt DNNF}}
\newcommand{\dDNNF}{{\tt d-DNNF}}
\newcommand{\sDNNF}{{\tt s-DNNF}}
\newcommand{\sdDNNF}{{\tt sd-DNNF}}
\newcommand{\sdNNF}{{\tt sd-NNF}}
\newcommand{\OBDD}{{\tt OBDD}}
\newcommand{\sOBDD}{{\tt s-OBDD}}
\newcommand{\MODS}{{\tt MODS}}
\newcommand{\OBDDo}{{\tt OBDD$_<$}}

\newcolumntype{C}[1]{>{\centering\let\newline\\\arraybackslash\hspace{0pt}}m{#1}}

\begin{document}
\graphicspath{{./figures/}}
\hyphenation{know-ledge Santos making efficiently Kimmig using semi-ring analy-sis}

\title{Algebraic Model Counting
}

\author{Angelika Kimmig         \and
       Guy Van den Broeck \and Luc De Raedt \\ \small{Department of Computer Science, KU Leuven}\\
       \small{Celestijnenlaan 200a -- box 2402, 3001 Heverlee, Belgium}\\
\small{\texttt{\{firstname.lastname\}@cs.kuleuven.be}}
}
\date{}

\maketitle

\begin{abstract}
Weighted model counting (WMC) is a well-known inference task on knowledge bases, used for probabilistic inference in graphical models.
We introduce algebraic model counting (AMC), a generalization of WMC to a semiring 
structure. We  show that  AMC generalizes many well-known tasks in a
variety of domains such as probabilistic inference, 
soft constraints and network and database analysis.  
Furthermore, we investigate AMC from a knowledge compilation perspective
and show that all AMC tasks can be evaluated using  \sdDNNF\ circuits. 
We  identify further characteristics of AMC instances that allow for the use of 
 even  more succinct circuits.
\end{abstract}

\section{Introduction}\label{sec:intro}
Today, some of the most efficient techniques for probabilistic
inference employ reductions to weighted model counting~(WMC) 
both for propositional and for relational
probabilistic models~\citep{Park02,Sang05,Darwiche09,Fierens11}.
 On the other hand, it is well-known that probabilistic
inference as well as many other tasks can be generalized to a sum of
products computation over models with suitable operators from a semiring structure.
This has led to
common inference algorithms for a variety of different inference
problems
in a variety of different fields, see for instance~\citet{Goodman99},
\citet{Eisner05}, \citet{Meseguer06}, \citet{Green2007},
\citet{bacchus2009solving}, \citet{Larrosa10}, \citet{baras2010path} and
\citet{Kimmig11}. The work presented here builds on these two lines of work.

As our first contribution, we introduce the task of
\emph{algebraic model counting (AMC)}.  AMC generalizes weighted
model counting to the semiring setting and supports various
types of labels,  including numerical ones as used in WMC,  but also sets,
polynomials, Boolean formulae, and many more. It thereby also  
generalizes many different tasks from a variety of different
fields.  
As our second contribution,  we investigate how to solve AMC problems
using knowledge compilation. As  AMC is defined in
terms of the set of models of a propositional logic theory, we can
exploit the succinctness results of 
the knowledge compilation map of~\citet{darwiche2002knowledge}.
We show that AMC can in general be  evaluated using \sdDNNF\ circuits,
which are more succinct than a direct representation of the set of models.  Furthermore, we
identify a number of characteristics of AMC tasks that allow for using
even more succinct types of circuits. 
Our results generalize well-known
results for satisfiability and model counting in circuits to broad classes of AMC tasks
and extend the task classification in algebraic Prolog~\citep{Kimmig11}
to more succinct types of circuits. 
As our third contribution,  we provide conditions under which AMC
generalizes 
semiring sums of products defined over derivations, that is,
sequences of possibly repeated variables, instead of over models.

This paper is organized as follows. We introduce algebraic model
counting in Section~\ref{sec:amc}. Section~\ref{sec:amc-kc} provides
task characteristics that allow for sound evaluation on specific
classes of circuits and shows how these generalize previous results. We
discuss future work and conclude in Section~\ref{sec:fw}.

\section{Algebraic Model Counting}\label{sec:amc}
We first define the task of algebraic model counting, provide some example
instances, and briefly discuss
its relationships to existing frameworks. The underlying mathematical structure is that of a
commutative semiring. 
\begin{definition}[(Commutative) Semiring]
A \emph{semiring} is a structure\linebreak[4] $(\mathcal{A}, \oplus, \otimes, 
e^{\oplus}, e^{\otimes})$, where \emph{addition}~$\oplus$ and
\emph{multiplication}~$\otimes$ are associative binary operations over
the set~$\mathcal{A}$, $\oplus$~is commutative, $\otimes$~distributes
over~$\oplus$, $e^{\oplus}\in\mathcal{A}$ is the neutral element of~$\oplus$, $e^{\otimes}\in\mathcal{A}$ that of~$\otimes$,
and for all $a\in \mathcal{A}$, $e^{\oplus}\otimes a = a \otimes
e^{\oplus} = e^{\oplus}$. In a \emph{commutative semiring}, $\otimes$~is
commutative as well.
\end{definition}
Algebraic model counting is now defined as follows:
\begin{definition}[AMC Problem]
Given
\begin{itemize}
  \item a \emph{propositional logic theory} $T$ over a set of
    variables $\mathcal{V}$, 
  \item a \emph{commutative semiring} $(\mathcal{A},\oplus,\otimes,   e^{\oplus},e^{\otimes})$, and 
 \item a \emph{labeling function} $\alpha : \mathcal{L} \rightarrow \mathcal{A}$, mapping literals $\mathcal{L}$ of the variables in $\mathcal{V}$ to elements of the semiring set $\mathcal{A}$, 
\end{itemize}
compute
\begin{equation} \label{eq:amc}
\operatorname{\mathbf{A}}(T)  = \bigoplus_{I\in \mathcal{M}(T)} \bigotimes_{l \in I} \alpha(l),
\end{equation}
where $\mathcal{M}(T)$ denotes the set of models of~$T$.
\end{definition}

\begin{sidewaystable}
\center
\begin{tabular}{|c||c|c|c|c|c|c|c||C{1.4cm}|}
\hline
task & $\mathcal{A}$ & $e^{\oplus}$ & $e^{\otimes}$ & $\oplus$ &
$\otimes$  & $\alpha(v)$ & $\alpha(\neg v)$ & ref\\\hline \hline
SAT & $\{true,false\}$& $false$& $true$& $\vee$& $\wedge$& $true$&
$true$& B,  BT, G, GK, K, L, M\\\hline
\#SAT & $\mathbb{N}$& $0$& $1$& $+$& $\cdot$& $ 1$&
$1$& B, G, GK, K, L\\\hline 
WMC & $\mathbb{R}_{\geq 0}$ & $0$ & $1$ & $+$ & $\cdot$ & $\in
\mathbb{R}_{\geq 0}$& $\in\mathbb{R}_{\geq 0}$ & \\\hline
PROB & $\mathbb{R}_{\geq 0}$ & $0$ & $1$ & $+$ & $\cdot$ & $\in[0,1]$&
$1-\alpha(v)$ & B, BT, E,  G, K\\\hline
SENS & $\mathbb{R}[\mathcal{V}]$& $0$& $1$& $+$&
$\cdot$& $v $ or $\in[0,1]$ &
$1-\alpha(v)$& K\\\hline
GRAD & $\mathbb{R}_{\geq 0}\times\mathbb{R}$& $(0,0)$& $(1,0)$ &
Eq.~\eqref{eq:gradoplus} &Eq.~\eqref{eq:gradotimes}  &
Eq.~\eqref{eq:gradalpha} & Eq.~\eqref{eq:gradoveralpha} & E, K\\\hline 
MPE & $\mathbb{R}_{\geq 0}$& $0$& $1$& $ \max$& $\cdot$& $\in[0,1]$&
$1-\alpha(v)$& B, BT, G, K, L, M\\\hline
S-PATH & $\mathbb{N}^{\infty}$&
$\infty$& $0$& $\min$& $+$& $\in\mathbb{N}$& $0$& BT,  GK, K\\\hline
W-PATH & $\mathbb{N}^{\infty}$& $0$&
$\infty$& $\max$& $\min$& $\in\mathbb{N}$& $\infty$ & BT\\\hline 
FUZZY & $[0,1]$ & $0$ & $1$ & $\max$ & $\min$ &  $\in [0,1]$ & $1$&
GK, M\\\hline
$k$WEIGHT & $\{0,\ldots, k\}$ & $k$ & $0$ & $\min$ & $+^k$ &  $\in
\{0,\ldots, k\}$ &  $\in \{0,\ldots, k\}$ & M\\\hline
\OBDDo & \OBDDo($\mathcal{V}$) & \OBDDo($0$) &  \OBDDo($1$) & $\vee$&
$\wedge$& \OBDDo($v$)& $\neg$\OBDDo($v$) & K\\\hline
 WHY & $\mathcal{P(V)}$ & $\emptyset$ & $\emptyset$ & $\cup$  &
$\cup$  & $  \{v\}$ & n/a &  GK\\\hline
$\mathcal{RA^+}$ & $\mathbb{N}[\mathcal{V}] $ & $0$ & $1$ & $+$ & $\cdot$
&  $  v$ & n/a &  GK\\\hline 
\end{tabular}
\caption{Examples of commutative semirings and labeling functions. The
  \textbf{WHY} and $\mathcal{RA^+}$
  provenance semirings apply to positive literals only. Reference key:
  B \citep{bacchus2009solving}, 
BT \citep{baras2010path}, 
E  \citep{Eisner02}, 
G \citep{Goodman99},
GK  \citep{Green2007}, 
K \citep{Kimmig11}, 
L \citep{Larrosa10},
 M \citep{Meseguer06}; more examples can be found in
  these references. }
\label{tab:ex}
\end{sidewaystable}

\begin{theorem}\label{th:instances}
AMC generalizes satisfiability (\textbf{SAT}), model
counting (\textbf{\#SAT}), weighted model counting (\textbf{WMC}),  
probabilistic inference (\textbf{PROB}), sensitivity analysis
(\textbf{SENS}), gradient (\textbf{GRAD}), probability of most likely states (\textbf{MPE}),
shortest (\textbf{S-PATH}) and widest (\textbf{W-PATH})  paths, fuzzy (\textbf{FUZZY})
and $k$-weighted (\textbf{$k$WEIGHT}) constraints, and \OBDDo\
construction.
\begin{proof}
By verification of the definitions in~Table~\ref{tab:ex}. 
\end{proof}
\end{theorem}
\textbf{SAT}, \textbf{\#SAT}, \textbf{WMC} and
\textbf{PROB} are well-known tasks that appear in many fields. 
\textbf{SENS} and \textbf{GRAD} allow one to investigate the effect
of parameter changes and to learn parameters in a
probabilistic setting, respectively. 
In \textbf{GRAD}, labels are tuples $(p_i,g_i)$ with  $p_i\in[0,1]$
the probability of $v_i$ and $g_i$ the gradient with respect to the
$k$th variable:
\begin{align}
\alpha(v_i) & =
  \left\{ \begin{array}{ll}
     (p_i,1) & \mbox{ if }i=k \\
    (p_i,0)& \mbox{ if }i\neq k
  \end{array} \right.\label{eq:gradalpha}\\
\alpha(\neg v_i) & =
  \left\{ \begin{array}{ll}
    (1-p_i, -1) & \mbox{ if }i=k \\
     (1-p_i, 0)& \mbox{ if }i\neq k
  \end{array} \right.\label{eq:gradoveralpha}\\
(a_1,a_2)\oplus (b_1,b_2) &= (a_1+b_1, a_2+b_2)\label{eq:gradoplus}\\
(a_1,a_2)\otimes (b_1,b_2) &= (a_1\cdot b_1, a_1\cdot b_2 + a_2\cdot b_1)\label{eq:gradotimes}
\end{align}
If the second element of the label denotes a cost, the \textbf{GRAD} semiring
calculates expected costs. 
\textbf{MPE}, \textbf{S-PATH},  \textbf{W-PATH},  \textbf{FUZZY}  and
\textbf{$k$WEIGHT} (with bounded addition~$+^k$) are examples of
optimization tasks. Finally, AMC can also
be used to construct a canonical representation of the set of models
in the form of an \OBDDo\ circuit, a popular data structure in many
fields of computer science. 
The last two tasks in the table originate from probabilistic databases under the positive 
relational algebra~$\mathcal{RA^+}$ and are not easily extended to
negative literals. Why-provenance (\textbf{WHY})
collects the set of identifiers of all tuples an answer depends on, whereas 
$\mathcal{RA^+}$-provenance constructs polynomials that also take into
account the number of times the tuples are used.
While all tasks listed in Table~\ref{tab:ex} are representative examples from the literature, cf.~the references given in
the table, this is by no means an exhaustive list of semirings
and labeling functions that can be used for AMC.

As these examples illustrate, 
the AMC task shares its basic structure with a number
of other tasks. 
The class of sum-of-products problems
generalizes factor graphs to the algebraic setting, but uses
discrete valued variables or factors as basic building
blocks~\citep{bacchus2009solving}. In this context, affine algebraic
decision diagrams~\citep{Sanner05} and AND/OR multi-valued decision
diagrams~\citep{Mateescu08} have been used for inference with real-valued semirings. The
restriction to two-valued variables allows us to directly compile AMC
tasks to propositional circuits without adding constraints on legal
variable assignments to the theory. In soft
constraint programming, additional constraints are
imposed on the semiring, which ensure that addition optimizes the
degree of constraint satisfaction~\citep{Meseguer06}. \cite{Wilson05}
provides an algorithm that compiles semiring-based systems into
semiring-labelled decision diagrams, which are closely related to
unordered binary decision diagrams, to compute valuations.  Semiring-induced
propositional logic labels clauses with semiring elements with a
weight associated to their falsification and is restricted to
semirings whose induced pre-order is partial~\citep{Larrosa10}. In algebraic Prolog
(aProbLog), a semiring-labeled logic program is
reduced to  AMC for inference~\citep{Kimmig11}. 

While AMC sums over models, other tasks sum over sequences
of possibly repeated variables. Examples include 
algebraic path
problems~\citep{baras2010path}, semiring
parsing~\citep{Goodman99}, provenance semirings for positive
relational algebra queries in databases~\citep{Green2007}, and
semiring-weighted dynamic programs~\citep{Eisner05}. We will discuss  the difference
between such derivation-based settings and AMC in 
more detail 
in Section~\ref{sec:amcapp}.

\section{AMC using Knowledge Compilation} \label{sec:amc-kc}
In their knowledge compilation map, \citet{darwiche2002knowledge}
provide an overview of succinctness relationships between various
types of propositional circuits. Furthermore, they show which reasoning tasks in
propositional logic, such as (weighted) model
counting (\textbf{\#SAT/WMC}) or satisfiability checking
(\textbf{SAT}), are evaluated on which circuits in time polynomial in
the size of the circuit.  Propositional circuits are often used as a
representation language in  weighted model counting and similar
tasks, including for instance probability calculation and
sensitivity analysis in probabilistic databases~\citep{Jha11,Kanagal11}
and  inference in algebraic
Prolog~\citep{Kimmig11}. In the following, we extend this approach to
AMC. We  first repeat the relevant knowledge compilation
concepts, closely following~\citet{darwiche2002knowledge}.

\begin{definition}[\NNF]
A sentence in \emph{negation normal form} (\NNF) over a set of propositional variables $\mathcal{V}$
is a rooted, directed acyclic graph where each leaf node is labeled
with true ($\top$), false ($\bot$), or a literal of a variable in~$\mathcal{V}$, and
each internal node with disjunction ($\vee$) or conjunction ($\wedge$). 
\end{definition}

\begin{definition}[Decomposability]
An \NNF\ is \emph{decomposable} if for each conjunction
node~$\bigwedge_{i=1}^n\phi_i$, no two children~$\phi_i$ and~$\phi_j$
share any variable. 
\end{definition}
\begin{definition}[Determinism]
An \NNF\ is \emph{deterministic} if for each disjunction
node~$\bigvee_{i=1}^n\phi_i$, each pair of different 
children~$\phi_i$ and~$\phi_j$ is logically inconsistent.
\end{definition}
\begin{definition}[Smoothness]
An \NNF\ is \emph{smooth} if for each disjunction
node $\bigvee_{i=1}^n\phi_i$, each child~$\phi_i$ mentions the
same set of variables.
\end{definition}
\DNNF, \dNNF, \sNNF, \sdNNF, \dDNNF, \sDNNF, and \sdDNNF\ are the
subsets of \NNF\ satisfying  (combinations of)
these properties, where \texttt{D} stands for decomposable, \texttt{d} for
deterministic, and \texttt{s} for smooth. 
\begin{figure}
\center
\subfloat[\sdDNNF]{\label{fig:circuits:left} \includegraphics[scale=0.35]{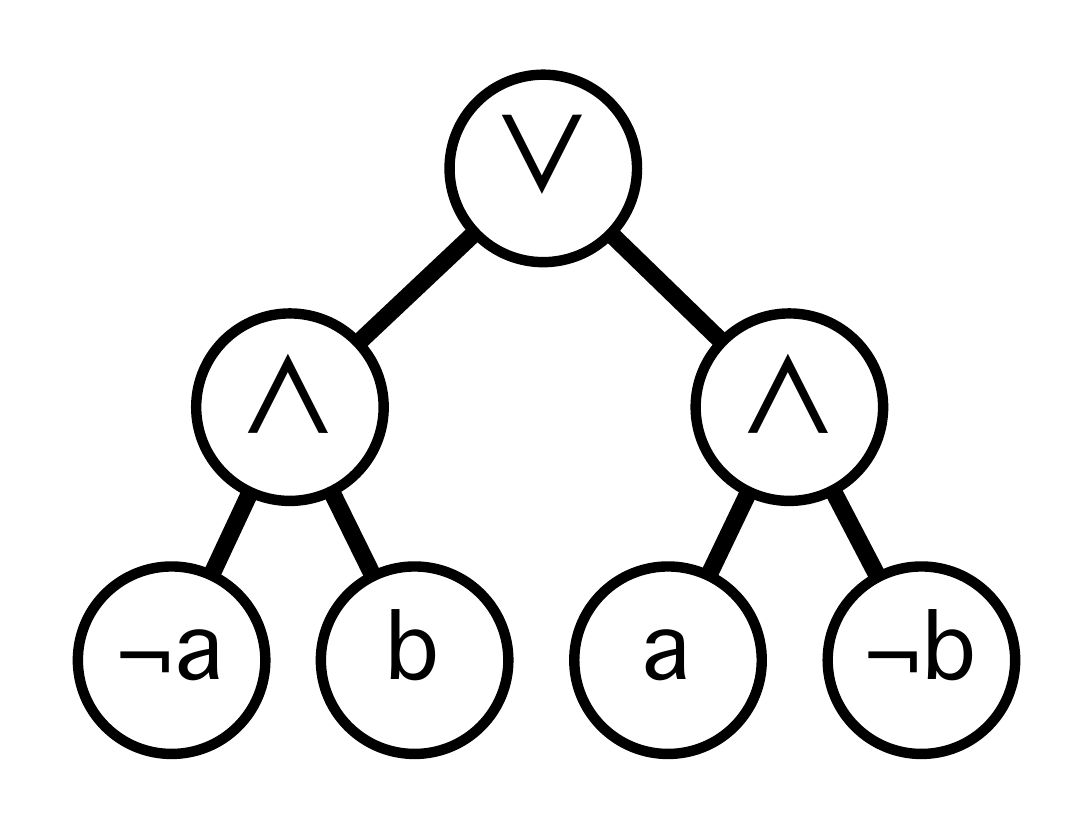}}
\subfloat[\NNF]{\label{fig:circuits:right} \includegraphics[scale=0.35]{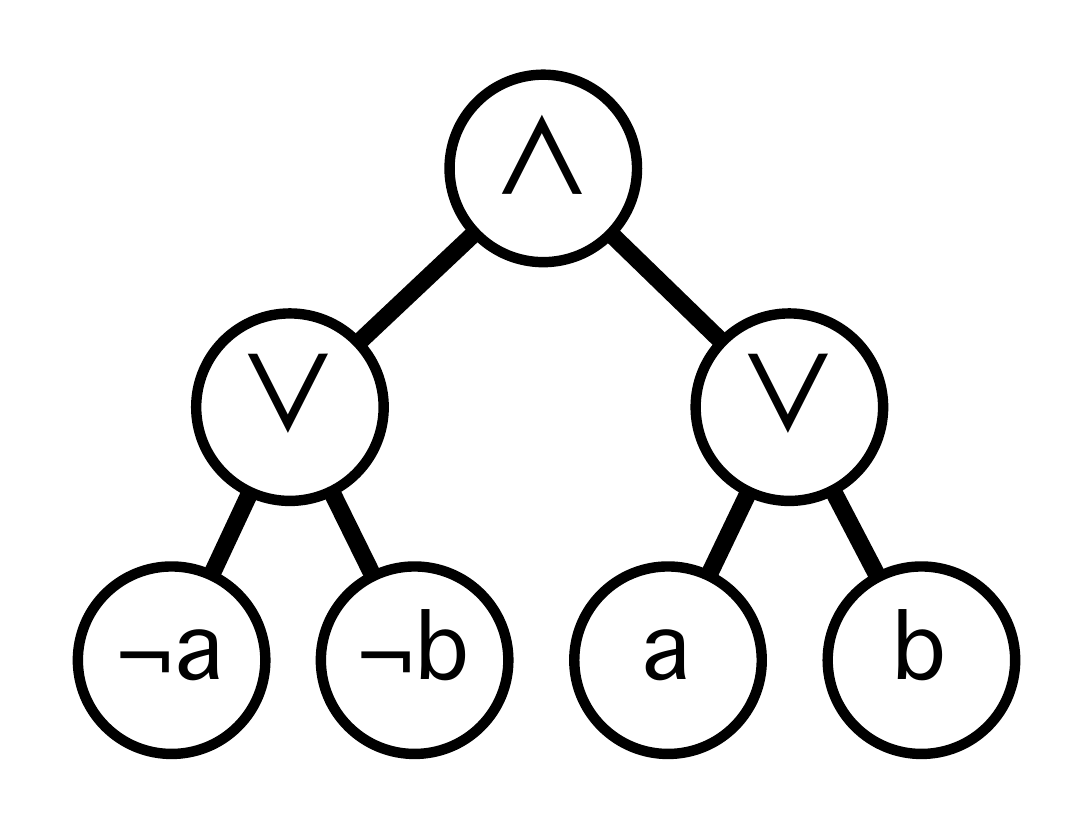}}
\caption{Example of an \sdDNNF\ and \NNF\ circuit.}
\label{fig:circuits}
\end{figure}
For instance, the circuit in Figure~\ref{fig:circuits:left} is in \sdDNNF,
while the one in Figure~\ref{fig:circuits:right} has none of the three properties.

The algebraic model count~$\operatorname{\mathbf{A}}(T)$ is defined as a summation over the set of
models~$\mathcal{M}(T)$ of a propositional theory~$T$, which corresponds to the \MODS\ language in
the knowledge compilation map. However, as this \MODS\ language is
exponentially less succinct than any other representation of~$T$
included in the map,
converting to \MODS\ in order to evaluate Equation~\eqref{eq:amc}
directly is undesirable.
 In the following, we therefore establish a connection between
 characteristics of AMC
 tasks and properties of the \NNF\ circuits they can be evaluated on,
 resulting in the classification scheme summarized in Table~\ref{tab:eval}.

\begin{table}
\center
\begin{tabular}{c||c|c||c|c}
& \multicolumn{2}{c||}{} & \multicolumn{2}{c}{idempotent and}\\ 
& \multicolumn{2}{c||}{general $\otimes$} & \multicolumn{2}{c}{consistency-pres.~$(\otimes,\alpha)$}\\ 
× & neutral & non-neutral & neutral & non-neutral \\ 
× & $(\oplus,\alpha)$ &$(\oplus,\alpha)$ & $(\oplus,\alpha)$ & $(\oplus,\alpha)$ \\ \hline \hline
idempotent $\oplus$ & \DNNF & \sDNNF & \NNF & \sNNF\\ 
& (Th.~\ref{th:DNNF}) & (Th.~\ref{th:sDNNF}) & (Th.~\ref{th:NNF}) & (Th.~\ref{th:NNF})  \\ \hline
non-idempotent $\oplus$ & \dDNNF & \sdDNNF & \dNNF & \sdNNF \\
 & (Th.~\ref{th:dDNNF}) & (Th.~\ref{th:sdDNNF}) & (Th.~\ref{th:NNF}) & (Th.~\ref{th:sdNNF})
\end{tabular}
\caption{Semiring characteristics and corresponding circuits that
  allow for sound AMC evaluation.}\label{tab:eval}
\end{table}

The key idea underlying \NNF\ evaluation is to perform a bottom-up
pass over the circuit, labeling each node with the value of the
subcircuit rooted at that node. For disjunction nodes, the values of
all their 
children are combined using~$\oplus$, for conjunction nodes using~$\otimes$.
\begin{definition}[\NNF\ Evaluation]
The function \textsc{Eval} specified in Algorithm~\ref{alg:eval} \emph{evaluates} an \NNF\ circuit for a commutative semiring $(\mathcal{A},\oplus,\otimes, e^{\oplus},e^{\otimes})$ and labeling function $\alpha$.
\end{definition}

\begin{algorithm}[t]
  \caption[\textsc{Label}]{Evaluating an \NNF\ circuit $N$ for a
    commutative semiring $(\mathcal{A},\oplus,\otimes,
    e^{\oplus},e^{\otimes})$ and labeling function $\alpha$.}
\label{alg:eval}
\begin{algorithmic}[1]
\FUNCTION{\textsc{Eval}($N,\oplus,\otimes,e^{\oplus},e^{\otimes},\alpha$)}
\STATE \textbf{if} $N$ is a true node $\top$ \textbf{then} return $e^{\otimes}$
\STATE \textbf{if} $N$ is a false node $\bot$ \textbf{then} return $e^{\oplus}$
\STATE \textbf{if} $N$ is a literal node $l$ \textbf{then} return $\alpha(l)$
\IF{$N$ is a disjunction $\bigvee_{i=1}^mN_i$} \STATE return $\bigoplus_{i=1}^m$ \textsc{Eval}($N_i,\oplus,\otimes,e^{\oplus},e^{\otimes},\alpha$) \ENDIF
\IF{$N$ is a conjunction $\bigwedge_{i=1}^mN_i$} \STATE return $\bigotimes_{i=1}^m$ \textsc{Eval}($N_i,\oplus,\otimes,e^{\oplus},e^{\otimes},\alpha$) \ENDIF
\ENDFUNCTION
\end{algorithmic}
\end{algorithm}

Consider for example \textbf{\#SAT} for the two circuits in
Figure~\ref{fig:circuits}, which both represent an exclusive OR of two
variables. Evaluation of the \sdDNNF\ in Figure~\ref{fig:circuits:left}, which in fact
is a \MODS\ representation, assigns label $1$ to each leaf, 
$1\cdot 1=1$ to each conjunction
node, and 
$1+1=2$ to the disjunction
node at the root and thus the entire circuit, which is correct. On the
other hand, evaluation on the general \NNF\   in
Figure~\ref{fig:circuits:right} assigns $1+1=2$ to each disjunction
node and $2\cdot 2=4$ to the conjunction node at the root. This
overestimation is due to models shared by the children of the same
disjunction node and variables shared by the children of the
conjunction node, as we will see in more detail in Section~\ref{sec:decomposable} and~\ref{sec:nondecomposable}.

\begin{definition}[Soundness]
Evaluating an \NNF\ representation~$N_T$ of a propositional theory~$T$
for a  semiring $(\mathcal{A},\oplus,\otimes,e^{\oplus},e^{\otimes})$
and labeling function~$\alpha$   is a
\emph{sound AMC computation} iff \textsc{Eval}$(N_T,\oplus,\otimes,e^{\oplus},e^{\otimes},\alpha) =\operatorname{\mathbf{A}}(T)$.
\end{definition}
In the following, we establish a general soundness result for AMC
evaluation on \sdDNNF\ circuits as
well as properties of AMC tasks that guarantee soundness for various
other subclasses of \NNF. Given soundness, we inherit the polynomial
complexity results of the knowledge compilation
map~\citep{darwiche2002knowledge} for 
semiring operators with unit cost.
Note however that there are semirings with more
expensive operators. For instance, labels in  \OBDDo\ may grow exponentially
in the circuit size.

\subsection{\sdDNNF\ Evaluation}
We show that AMC evaluation is sound on \sdDNNF\ circuits.
As these are strictly more succinct than \MODS\ representations, they allow for
more efficient inference.
\begin{theorem}[\sdDNNF\ Evaluation]\label{th:sdDNNF}
  Evaluating an \sdDNNF\ representation of the propositional theory~$T$ is a sound AMC computation.
  \begin{proof}
  We show that \textsc{Eval}($N_T,\oplus,\otimes,e^{\oplus},e^{\otimes},\alpha$) for an \sdDNNF\
  representation $N_T$ of the theory $T$  computes
  $\operatorname{\mathbf{A}}(T)$ with respect to all variables
  in~$N_T$:
\begin{enumerate}
\item Line 2: 
  $\operatorname{\mathbf{A}}(\top) =
  \bigoplus_{I\in\{\emptyset\}}\bigotimes_{l\in I}\alpha(l) = e^{\otimes}$
 \item Line 3: 
  $\operatorname{\mathbf{A}}(\bot) =
  \bigoplus_{I\in \{\}}\bigotimes_{l\in I}\alpha(l) = e^{\oplus}$
\item Line 4: 
  $\operatorname{\mathbf{A}}(l) =
  \bigoplus_{I\in\{\{l\}\}}\bigotimes_{k\in I}\alpha(k) = \alpha(l)$
\end{enumerate}
Due to associativity and commutativity of the semiring
  operators, operands of each summation and each multiplication can be
  evaluated in arbitrary order. We therefore restrict ourselves to
  binary disjunction and conjunction nodes here. 
 Given sound evaluation for subcircuits~$\phi_i$
  with variables~$\mathcal{V}_i$ and models~$\mathcal{M}_i$ with
  respect to these variables, we obtain:
\begin{enumerate}
\setcounter{enumi}{3}
\item Lines 5-6:  
Disjunction node $\phi_1\vee\phi_2$: 
As  $\mathcal{V}=\mathcal{V}_1\cup\mathcal{V}_2 = \mathcal{V}_i$
due to smoothness, we obtain $\mathcal{M}(\phi_1\vee\phi_2) = 
\mathcal{M}_1\cup\mathcal{M}_2$, which is a disjoint union due to determinism. Therefore,
$\operatorname{\mathbf{A}}(\phi_1)\oplus
\operatorname{\mathbf{A}}(\phi_2) =
\bigoplus_{i=1,2}\bigoplus_{\mathcal{M}_i}\bigotimes_{l\in I}\alpha(l)
= \bigoplus_{\mathcal{M}_1\cup \mathcal{M}_2}\bigotimes_{l\in
  I}\alpha(l) = \operatorname{\mathbf{A}}(\phi_1\vee\phi_2)$.
\item Lines 7-8: 
Conjunction node $\phi_1\wedge\phi_2$: As
$\mathcal{V}_1\cap\mathcal{V}_2=\emptyset$ due to decomposability,
the set $\mathcal{M}(\phi_1\wedge\phi_2)$ of models of the conjunction
is simply the set of all unions of models of its parts. Together with distributivity, we get $\operatorname{\mathbf{A}}(\phi_1)\otimes
\operatorname{\mathbf{A}}(\phi_2) =
\bigotimes_{i=1,2}\bigoplus_{\mathcal{M}_i}\bigotimes_{l\in
  I}\alpha(l) = \bigoplus_{\mathcal{M}(\phi_1\wedge\phi_2)}\bigotimes_{l\in
  I}\alpha(l) = \operatorname{\mathbf{A}}(\phi_1\wedge\phi_2)$.
\end{enumerate}\end{proof}
\end{theorem}

Clearly, the soundness of AMC evaluation on  \sdDNNF\ depends on all three
properties of this subclass of \NNF. On the other hand, circuits without these properties may be exponentially smaller and thus allow
for more efficient inference. 
In the following, we therefore analyze evaluation in the absence of these properties, which allows us to identify characteristics of the semiring and labeling function that ensure sound evaluation on the corresponding classes
of circuits.

\subsection{Evaluation on other Decomposable Circuits}\label{sec:decomposable}
If a circuit is not deterministic, children of a disjunction node may have
common models, in which case evaluation sums over such shared models
multiple times. For instance, consider the circuit in
Figure~\ref{fig:circuits:right} with \textbf{PROB}, $\alpha(a)=0.6$ and
$\alpha(b)=0.3$. Evaluation on this circuit results in $0.6+0.3=0.9$ for
the right disjunction node, while $\operatorname{\mathbf{A}}(a \vee b) = 0.6 \cdot 0.3 +  0.4 \cdot 0.3 + 0.6 \cdot 0.7
= 0.72$.

\begin{definition}[Idempotent Operator]
  A binary operator~$\odot$ over a set~$\mathcal{A}$ is \emph{idempotent} iff $\forall a \in \mathcal{A}: a\odot a=a$.
 \end{definition}

\begin{theorem}[\sDNNF\ Evaluation]\label{th:sDNNF}
 Evaluating an \sDNNF\ representation of the propositional theory~$T$ for a semiring with idempotent~$\oplus$ is a sound AMC computation.
 \begin{proof}
Reconsider point (4) of the proof of Theorem~\ref{th:sdDNNF}. 
With smoothness, but without determinism,
$\mathcal{M}_1(\phi_1)\cup\mathcal{M}_2(\phi_2)$ is no longer a union
of disjoint sets, and $\operatorname{\mathbf{A}}(\phi_1)\oplus
\operatorname{\mathbf{A}}(\phi_2) =
\bigoplus_{i=1,2}\bigoplus_{\mathcal{M}_i}\bigotimes_{l\in
  I}\alpha(l)$ sums over the
models in $\mathcal{M}_1(\phi_1)\cap\mathcal{M}_2(\phi_2)$ twice. 
Due to associativity and commutativity,
   this is sound for idempotent~$\oplus$.
  \end{proof}
\end{theorem}

If a circuit is not smooth, the children of a disjunction node may use
different sets of variables. 
Each model of a child node corresponds to a set of models for the full
set of variables, but evaluation on a non-smooth circuit ignores the labels of unmentioned
variables. For instance, consider the circuit in
Figure~\ref{fig:circuits:right} with \textbf{MPE}, $\alpha(a)=0.6$ and
$\alpha(b)=0.3$. Evaluating the right disjunction node of this circuit results in $\max(0.6,0.3)=0.6$, while $\operatorname{\mathbf{A}}(a \vee b) = \max(0.6 \cdot 0.3, 0.4 \cdot 0.3, 0.6 \cdot 0.7)
= 0.42$.

\begin{definition}[Neutral $(\oplus,\alpha)$]
  A semiring addition and labeling function pair~$(\oplus,\alpha)$ is 
  \emph{neutral} iff $\forall v \in \mathcal{V}: \alpha(v) \oplus \alpha(\neg v) =  e^{\otimes}$.
\end{definition}

\begin{theorem}[\dDNNF\ Evaluation] \label{th:dDNNF}
  Evaluating a \dDNNF\ representation of the propositional theory~$T$
  for a semiring and labeling function with  neutral~$(\oplus,\alpha)$
  is a sound AMC computation.
  \begin{proof}
For lines 5-6 (point (4) of the proof of Theorem~\ref{th:sdDNNF}) to
be sound, the sum of the  AMCs computed by the children over their
sets of variables $V_i$ has to be equal to the AMC of the entire disjunction
over the full set of variables $V_1 \cup V_2$. Given the child AMC
$\mathbf{A}_{V_i}(\phi_i)$, adding a variable $v$ to $V_i$ replaces
each model $I$ of $\phi_i$ by two models $I^+=I\cup \{v\}$ and $I^-
= I\cup \{\neg v\}$.
Due to distributivity, commutativity and the neutral sum property, the algebraic sum of these two models equals the AMC of the original model:
\begin{align*}
\mathbf{A}_{V_i \cup \{v\}}(I) &= \mathbf{A}_{V_i \cup \{v\}}(I^+) \oplus \mathbf{A}_{V_i \cup \{v\}}(I^-) \\
&= (\alpha(v)\oplus\alpha(\neg v)) \otimes \bigotimes_{l\in I}\alpha(l) \\
&= \bigotimes_{l\in I}\alpha(l)= \mathbf{A}_{V_i}(I)
\end{align*}
Evaluation therefore computes $\mathbf{A}_{V_1}(\phi_1)\oplus
\mathbf{A}_{V_2}(\phi_2)=\mathbf{A}_{V_1\cup V_2}(\phi_1)\oplus
\mathbf{A}_{V_1\cup V_2}(\phi_2)$, which due to determinism is equal
to $\mathbf{A}_{V_1\cup V_2}(\phi_1\vee \phi_2)$.
  \end{proof}
\end{theorem}

Note that from a practical point of view, non-neutral
$(\oplus,\alpha)$ does not influence the tractability of inference, as
any \NNF\ can be smoothed 
in polytime preserving determinism and decomposability~\citep{darwiche2002knowledge}.

The previous two results can directly be combined for \DNNF\ circuits
that are neither smooth nor deterministic.
\begin{theorem}[\DNNF\ Evaluation] \label{th:DNNF}
  Evaluating a \DNNF\ representation of the propositional theory~$T$
  for a semiring and labeling function with idempotent and  neutral~$(\oplus,\alpha)$
  is a sound AMC computation.
\begin{proof}
Reconsider point (4) of the proof of Theorem~\ref{th:sdDNNF}.
Due to neutral~$(\oplus,\alpha)$, values for all children of a disjunction node are
correct with respect to the full set of variables
(Theorem~\ref{th:dDNNF}). Due to idempotent~$\oplus$, multiple occurrences  of the same model
do not influence the result (Theorem~\ref{th:sDNNF}). 
  \end{proof}
\end{theorem}

This completes the left part of Table~\ref{tab:eval}, where no
conditions are imposed on semiring multiplication.

\subsection{Evaluation on Non-Decomposable Circuits}\label{sec:nondecomposable}
If a circuit is not
decomposable, the children of a conjunction node may share
variables. In this case, 
simply combining all pairs of their models may
produce sets of literals that are not models, because they either contain
contradicting literals, or several copies of the
same literal, which results in erroneous extra multiplications. For
instance, in Figure~\ref{fig:circuits:right}, $\{\neg
a,b\}$ is a model of both disjunction nodes, and $\{a,b\}$ of the
right one only. The conjunction node sums among others the products
$\alpha(\neg a)\otimes\alpha(b)\otimes\alpha(a)\otimes\alpha(b)$,
which does not correspond to a model, and 
$\alpha(\neg a)\otimes\alpha(b)\otimes\alpha(\neg a)\otimes\alpha(b)$,
where labels of both literals are multiplied twice.
\begin{definition}[Consistency-Preserving $(\otimes,\alpha)$]\label{def:cp}
  A semiring multiplication and labeling function pair~$(\otimes,\alpha)$ is 
  \emph{consistency-preserving} iff $\forall v \in
  \mathcal{V}: \alpha(v) \otimes \alpha(\neg v) = e^{\oplus}$. 
\end{definition}

\begin{theorem}[\sdNNF\ Evaluation]\label{th:sdNNF}
 Evaluating an \sdNNF\ representation of the propositional theory~$T$
 for a semiring and labeling function with  idempotent and
  consistency-preserving $(\otimes,\alpha)$ is a sound AMC computation.
 \begin{proof}
 Reconsider point (5) of the proof of Theorem~\ref{th:sdDNNF}.
The set of models of  $\phi_1\wedge\phi_2$ contains exactly all
pairwise combinations of models of its parts that agree on all shared variables.
The circuit evaluates the AMC of  $\phi_1\wedge\phi_2$ as
$\operatorname{\mathbf{A}}(\phi_1)\otimes
\operatorname{\mathbf{A}}(\phi_2)$, which due to distributivity is the
sum over all pairwise combinations of models. Without
decomposability, each such combination $I$ contains two literals for each $v\in
\mathcal{V}_1\cap\mathcal{V}_2$. As $\otimes$ is associative, commutative
and idempotent, repeated occurrences of a literal $l$ in
$\otimes_{i\in I}\alpha(i)$ do not affect the result. If $\{l,\neg
l\}\subseteq I$, $\otimes_{i\in I}\alpha(i)$ includes a multiplication by
$\alpha(l)\otimes\alpha(\neg l) = e^{\oplus}$, which in a semiring
means $\otimes_{i\in I}\alpha(i)  = e^{\oplus}$. Such inconsistent $I$ thus  do not
contribute to the semiring sum.
 \end{proof}
\end{theorem}
Theorem~\ref{th:sdNNF} affects only conjunction nodes, whereas
Theorems~\ref{th:sDNNF}, \ref{th:dDNNF} and \ref{th:DNNF} only affect
disjunction nodes. Their combination thus extends our results to 
non-decomposable circuits that do not satisfy (one of) the other two properties either:
\begin{theorem}[\sNNF, \dNNF, and \NNF\ Evaluations]\label{th:NNF}
 For a semiring and labeling function with idempotent and
  consistency preserving~$(\otimes,\alpha)$, evaluating the following representation of the propositional theory~$T$
  is a sound AMC computation:
\begin{itemize}
\item \sNNF\ if $\oplus$ is idempotent
\item \dNNF\ if $(\oplus,\alpha)$ is neutral
\item \NNF\ if $(\oplus,\alpha)$ is idempotent and neutral
\end{itemize}
\end{theorem}

This completes the right part of Table~\ref{tab:eval}, where using non-decomposable circuits is possible as a consequence of
restrictions on semiring multiplication. 
Given a new AMC instance, this table allows one  to immediately
choose the appropriate type of circuit for efficient
evaluation. For the examples discussed here, cf.~Table~\ref{tab:ex}, these are: \NNF\
for \OBDDo, \DNNF\ for  \textbf{SAT}, \textbf{S-PATH}, \textbf{W-PATH}
and \textbf{FUZZY}, \dDNNF\ for \textbf{PROB},
\textbf{SENS} and \textbf{GRAD}, \sDNNF\ for \textbf{MPE} and 
\textbf{$k$WEIGHT}, and \sdDNNF\ for \textbf{\#SAT} and \textbf{WMC}.

\subsection{Discussion}\label{sec:related} 
The results summarized in Table~\ref{tab:eval} generalize the
complexity results for \textbf{SAT} and \textbf{\#SAT}, provide more succinct types of
circuits for inference in algebraic Prolog, and show
that all circuits that are practically relevant for AMC are
well-studied in the knowledge compilation map. We now address these
points in more detail.  

First, \citet{darwiche2002knowledge} show that \textbf{SAT} can be
evaluated in polynomial time on \DNNF, while \textbf{\#SAT} can be evaluated
in polynomial time on \dDNNF, as smoothing is possible in polynomial
time. 
Our Theorem~\ref{th:DNNF} generalizes
soundness of \DNNF\ evaluation to all semirings and labeling functions
with idempotent and neutral $(\oplus,\alpha)$, which includes
\textbf{SAT}, while 
our Theorem~\ref{th:sdDNNF} generalizes soundness of \sdDNNF\ evaluation
to all semirings and labeling functions, including those with non-idempotent, non-neutral $(\oplus,\alpha)$,
such as \textbf{\#SAT}. 
As discussed above, the complexity of evaluation is polynomial if each 
semiring operation has unit cost.

Second, \citet{Kimmig11} reduce 
 inference in algebraic Prolog (aProbLog) to AMC evaluation on 
disjunctive normal form
(\DNF). For non-idempotent addition, the \DNF\ is compiled into an  ordered binary decision diagram
(\OBDD) if its conjunctions are not mutually exclusive. For non-neutral~$(\oplus,\alpha)$, circuits
are smoothed before evaluation. 
This results  in the following settings:  

\begin{center}
\begin{tabular}{|c|c|c|}
\hline
& neutral~$(\oplus,\alpha)$ & non-neutral~$(\oplus,\alpha)$  \\\hline
idempotent~$\oplus$ & \DNF & \sDNF \\\hline
non-idempotent~$\oplus$ & \dDNF & \sdDNF\\
 & \OBDD & \sOBDD  \\\hline
\end{tabular}
\end{center}
Table~\ref{tab:eval} uses the same characteristics of semiring operators and labeling
function, but does not assume a \DNF\ as starting point and thus also
does not rely on properties of such a \DNF. Its left half generalizes
the aProbLog scheme to more succinct superclasses, namely (\texttt{s-})\DNNF\ instead of (\texttt{s-})\DNF,
and (\texttt{s})\dDNNF\ instead of (\texttt{s})\dDNF\ or (\texttt{s-})\OBDD.

Third, we observe that while there are interesting inference tasks
that allow for sound evaluation on the more succinct class of \DNNF\
instead of the general \sdDNNF\ evaluation, the conditions for sound
evaluation on non-decomposable circuits are too strict in most
practical cases. 
This is in line with the knowledge compilation map,
which excludes non-decomposable circuits (with the exception of the
most general class \NNF) as they do not support any of the 
studied tasks in polytime~\citep{darwiche2002knowledge}.

\subsection{AMC and Algebraic Derivation Counting}\label{sec:amcapp}
While AMC is a sum over models, many other semiring-based tasks
require a 
sum over derivations, that is, sequences of possibly repeated variables. Examples include 
algebraic path
problems~\citep{baras2010path}, semiring
parsing~\citep{Goodman99}, provenance semirings for positive
relational algebra queries in databases~\citep{Green2007}, and
semiring-weighted dynamic programs~\citep{Eisner05}. 
We refer to this type of task as \emph{algebraic derivation
  counting} (ADC). While AMC and ADC appear very similar, they cannot
easily be exchanged, as we illustrate next. We restrict the
discussion to finite ADC.

\begin{figure}
\center
\subfloat[Path problem]{\label{fig:adc:left} \includegraphics[scale=0.35]{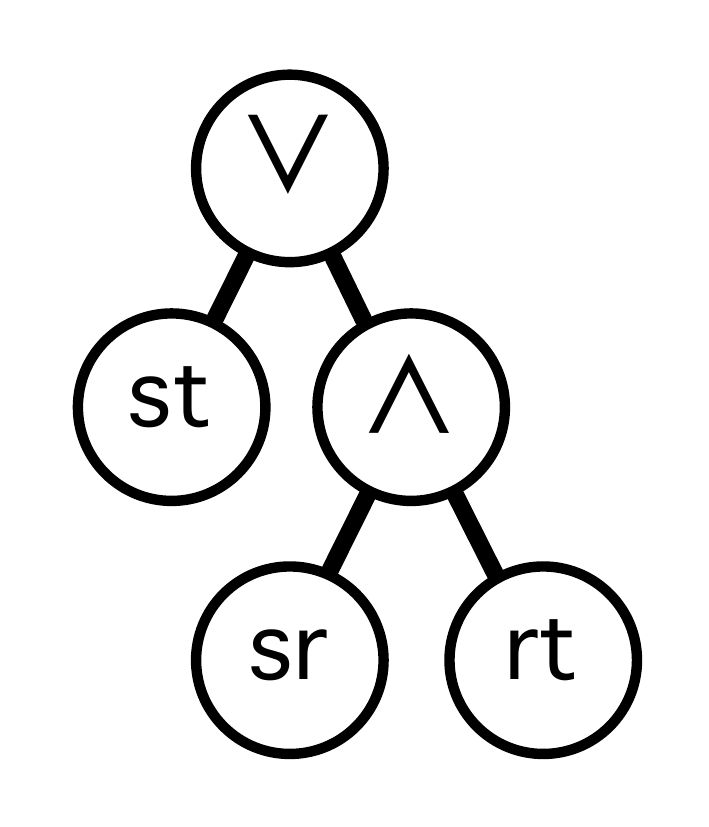} }
\subfloat[Semiring grammar]{\label{fig:adc:right} \includegraphics[scale=0.35]{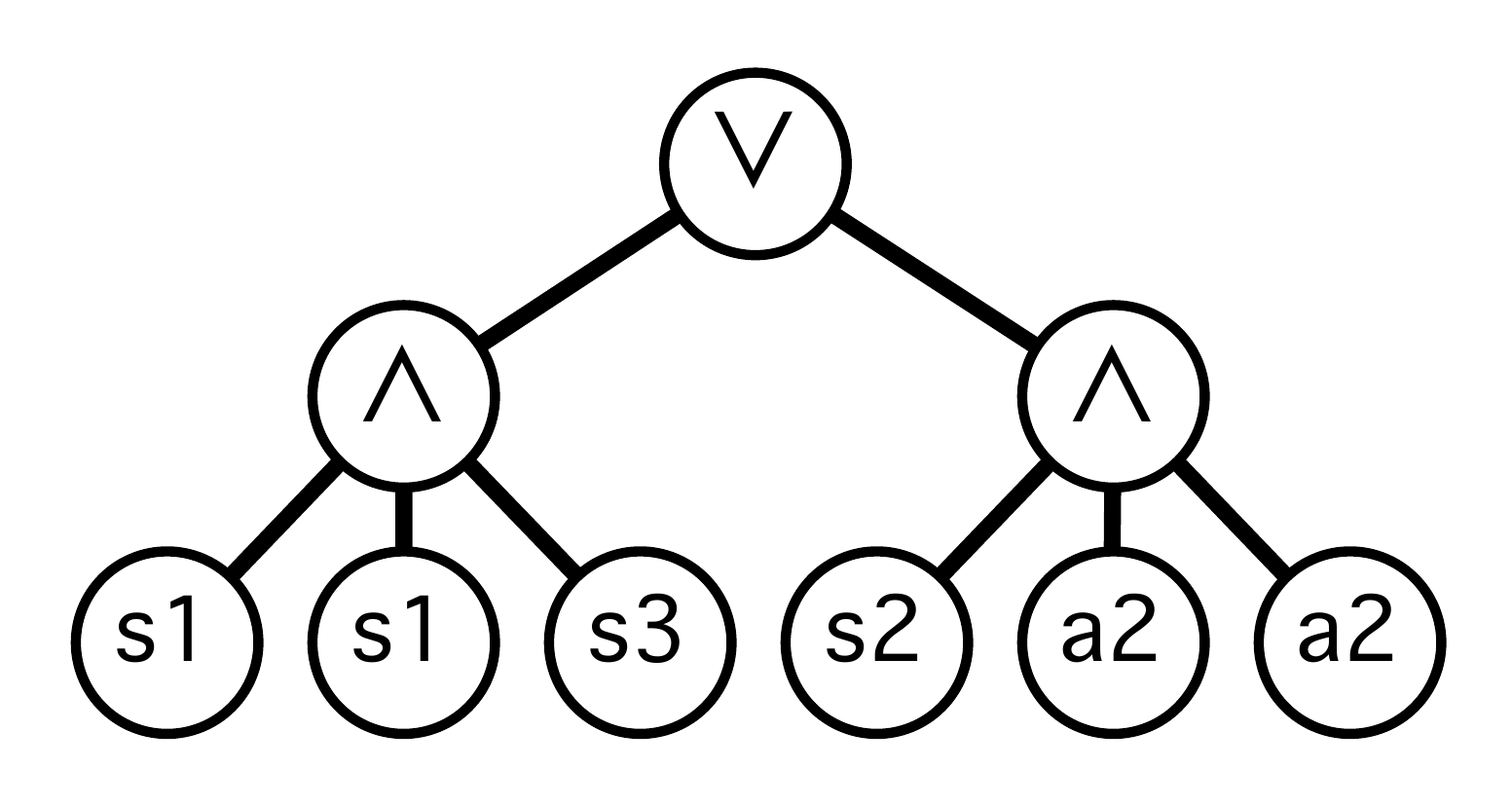} }
\caption{\NNF\ circuits for an algebraic path problem and a
  semiring grammar. See Section~\ref{sec:amcapp} for details.}
\label{fig:adc}
\end{figure}

Where AMC is based on a \MODS\ representation, that is, a smooth,
deterministic \DNF, ADC is based on a type of \NNF\ that is not
included in the knowledge compilation map: a disjunction of
conjunctions  of possibly repeated variables. Because of the repeated
variables, this is not a \DNF\ in general.   Figure~\ref{fig:adc}
shows two examples of such circuits. The \NNF\ in Figure~\ref{fig:adc:left} encodes the two paths between
nodes $s$ and $t$ in a graph with three edges $(s,t)$, $(s,r)$ and
$(r,t)$. It  is decomposable, as the conjunction does not repeat
variables, but neither smooth nor deterministic and thus a \DNNF. The \NNF\ in Figure~\ref{fig:adc:right} encodes the two derivations of $aa$ in a context-free grammar
with rules $S\rightarrow aS ~|~ AA ~|~ \epsilon$ and $A\rightarrow AA
~|~ a$. Variable $x_i$ denotes an application of 
the $i$th production rule for a non-terminal~$X$, that is, the right
conjunction in the circuit encodes the derivation $S\rightarrow AA
\rightarrow aA \rightarrow aa$. 
The circuit  is neither decomposable nor smooth, and  also not
deterministic, even though the derivations in the underlying grammar
setting are mutually exclusive (cf.~below). 

In order to represent an ADC task as an AMC task, the
labeling function~$\alpha$ needs to be extended to negative
literals. Together with the semiring, such an extension determines one of the
settings of Table~\ref{tab:eval}. If the ADC's \NNF\ belongs to the class of circuits corresponding to this
setting, the ADC and AMC tasks coincide. Clearly, this is always the
case for commutative semirings and labeling functions with idempotent, consistency-preserving
$(\otimes,\alpha)$ and idempotent, neutral $(\oplus,\alpha)$ such as 
for instance \OBDDo. However, as noted above, such tasks are rare.

For an ADC instance with idempotent addition defined on a \DNNF, that
is, without repeated variables in conjunctions, an
equivalent AMC instance can be constructed if $\alpha$
can be extended such that $(\oplus,\alpha)$ is neutral. For instance, consider the
circuit in Figure~\ref{fig:adc:left} in a shortest path
setting. By setting $\alpha(\neg v)=0$ for all variables $v$, the \textbf{S-PATH} semiring in Table~\ref{tab:ex} provides an
equivalent AMC task.
 
As the \NNF\ circuits of ADC do not contain negation, they are not 
deterministic. However, the underlying
tasks often impose additional constraints on their variables. For
instance, in the grammar example in Figure~\ref{fig:adc:right}, the leftmost
children $s1$ and $s2$ of the
two conjunction nodes are in fact mutually exclusive, as only one of
the right hand sides for $S$ can be chosen as first step in a
derivation. Such an $n$-ary variable can be encoded using $n$
binary variables by adding corresponding constraints to the theory
that restrict legal value assignments. For some semirings,
for instance \textbf{PROB}, it is sufficient to adapt the labels of
these variables without adding constraints. It is an open question under which general
conditions such label transformations are possible.

Conversely, every AMC task can be trivially represented as an ADC task
by introducing a derivation for each model. However, this is clearly
not desirable from a complexity point of view. 
\citet{baras2010path} provide an ADC encoding of network
reliability under probabilistic edge failure, that is, the \textbf{PROB} AMC
task for a positive propositional formula. They essentially modify multiplication to filter repeated literals,
and addition to subtract shared models of its operands, which 
drastically 
increases complexity of these operations. 
Again, it is an open question under which general
conditions such transformations are possible. 

\section{Conclusions and Future Work}\label{sec:fw}
We have introduced the task of algebraic model counting, which
generalizes weighted model counting to a semiring setting and thus
to various types of labels, including numerical ones as used in WMC,  but also sets,
polynomials, or Boolean formulae. We have shown that evaluating
AMC  is sound on \sdDNNF\ circuits, which are known
to be more succinct than the \MODS\ language used in its definition. Furthermore, we have provided characteristics of AMC tasks that
guarantee sound evaluation on more succinct classes of circuits.
This classification not only provides a means of directly choosing a circuit type
that allows for efficient inference given a new AMC task, but also 
generalizes a number of known results
and provides a framework to map restricted types of algebraic
derivation 
counts onto AMC tasks. 

Given the results presented here, it is worth investigating which other algebraic representations can be reduced to
algebraic model counting. 
Another line of future work concerns the introduction of additional
operators that would make it possible to express additional tasks, for
instance, partial MAP, which requires a maximization
operator in addition to summation and multiplication.

\paragraph{Acknowledgements} A.~Kimmig and G.~Van den Broeck are
supported by the Research Foundation Flanders (FWO Vlaanderen).


\end{document}